\documentclass[twocolumn,amsmath,amssymb]{revtex4}

\usepackage{graphicx}
\usepackage{dcolumn}
\usepackage{bm}

\begin{document}

\preprint{lateral qdm time-resolved}

\title{Influence of the Charge Carrier Tunneling Processes on the Recombination Dynamics in Single Lateral Quantum Dot Molecules}

\author{C.~ Hermannst\"adter$^{1}$}
\author{G.~J.~Beirne$^{1,}$}
 \altaffiliation{now at: The Cavendish Laboratory, University of Cambridge, J.~J.~Thomson Ave., Cambridge, CB3~0HE, UK}
\author{M.~Witzany$^1$}
\author{M.~Heldmaier$^1$}
\author{J.~Peng$^2$}
\author{G.~Bester$^2$}
\author{L.~Wang$^{2,3}$}
\author{A.~Rastelli$^3$}
\author{O.~G.~Schmidt$^3$}
\author{P.~Michler$^1$}
\affiliation{$^1$IHFG, Universit\"at Stuttgart, Allmandring 3, 70569 Stuttgart, Germany}
\affiliation{$^2$Max-Planck-Institut f\"ur Festk\"orperforschung, Heisenbergstr. 1, 70569 Stuttgart, Germany}
\affiliation{$^3$Institut f\"ur Integrative Nanowissenschaften, IFW Dresden, Helmholtzstr. 20, 01069 Dresden, Germany}


\begin{abstract}

We report on the charge carrier dynamics in single lateral quantum dot molecules and the effect of an applied electric field on the molecular states. Controllable electron tunneling manifests itself in a deviation from the typical excitonic decay behavior which is strongly influenced by the tuning electric field and inter-molecular Coulomb energies. A rate equation model is developed to gain more insight into the charge transfer and tunneling mechanisms. Non-resonant (phonon-mediated) electron tunneling which changes the molecular exciton character from direct to indirect, and vice versa, is found to be the dominant tunable decay mechanism of excitons besides radiative recombination.

\end{abstract}

\maketitle

The charge carrier configuration and dynamics in coupled quantum dot (QD) systems are the essential properties that need to be understood in order to gain the ability to coherently manipulate the coupling in the system using external electric, magnetic or light fields. This degree of control over the QD system represents an essential step toward the realization of quantum gates. Over the past number of years, optically addressable self-assembled semiconductor single QDs have been presented as sources for triggered single-photons and polarization entangled photon pairs \cite{Michl0,Michl1,Steve1,Akopi1,Hafen1}, and first quantum gates have been demonstrated \cite{Bonad1,Li1}. Such QDs can be assembled to larger molecular structures by vertically stacking them along the growth direction \cite{Solom1,Bayer1,Krenn1,Ortne1,Stina1,Krenn2,Schei1} and laterally arranging them \cite{Schmi1,Songm1,Beirn1}. A recent demonstration on a vertical QD molecule (QDM) has shown conditional quantum dynamics with one QD state being controlled via the other one \cite{Roble1}. 
The static properties of different types of QDMs, such as their coupling mechanisms and electronic structure, as well as, emitted photon characteristics have been extensively experimentally studied and theoretically described \cite{Beste1,Krenn1,Ortne1,Stina1,Wang1,Peng.Hermannstaedter.ea:2009}. A detailed dynamical analysis of the coupling in lateral QDMs using time-resolved spectroscopic methods, however, has not yet been done. As previously reported in Refs.~\cite{Beirn1,Wang1,Peng.Hermannstaedter.ea:2009}, the dominant coupling mechanism in lateral double-dots is electronic tunneling, which strongly depends on the charge carriers' effective masses, the excitonic binding energies and the potential landscape. It is therefore of particular interest to study the dependence of the tunnel dynamics on these parameters, especially considering the long-term objective of gaining the ability to control them in a deterministic way.
In this report we introduce our results obtained for experimental and theoretical examinations of the charge carrier and exciton dynamics of laterally coupled QDs which highlight the significant difference to single-dots and the important impact of a manipulating electric field.

\begin{figure}[bt]
	\centering
		\includegraphics[width=0.9\columnwidth]{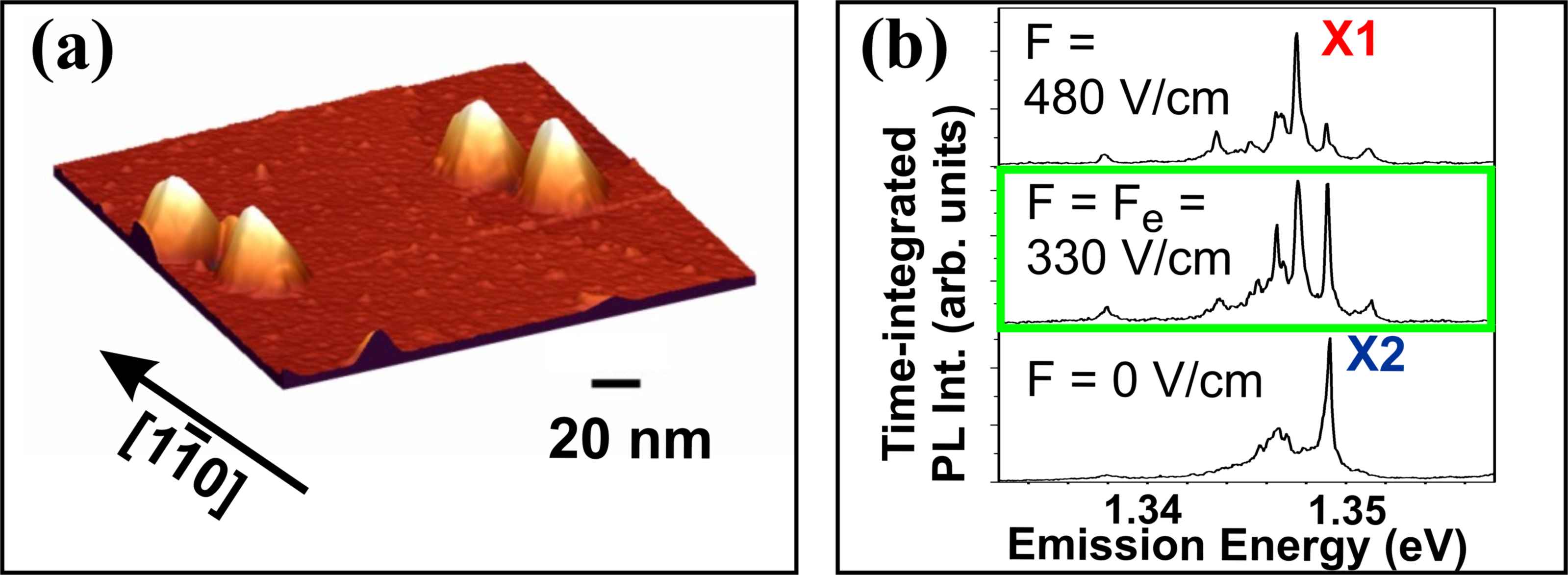}
	\caption{\textbf{(a)} AFM image showing two uncapped lateral QDMs. 
	\textbf{(b)} Time-integrated PL spectra of a single QDM for the different alignment situations at three different electric field strengths, $F=F_e=330$~V/cm (\emph{alignment case}), $F=480$~V/cm and $F=0~V/cm$ (\emph{misalignment cases}).
}
	\label{fig:Fig1}
\end{figure}

\begin{figure}[bt]
	\centering
		\includegraphics[width=0.9\columnwidth]{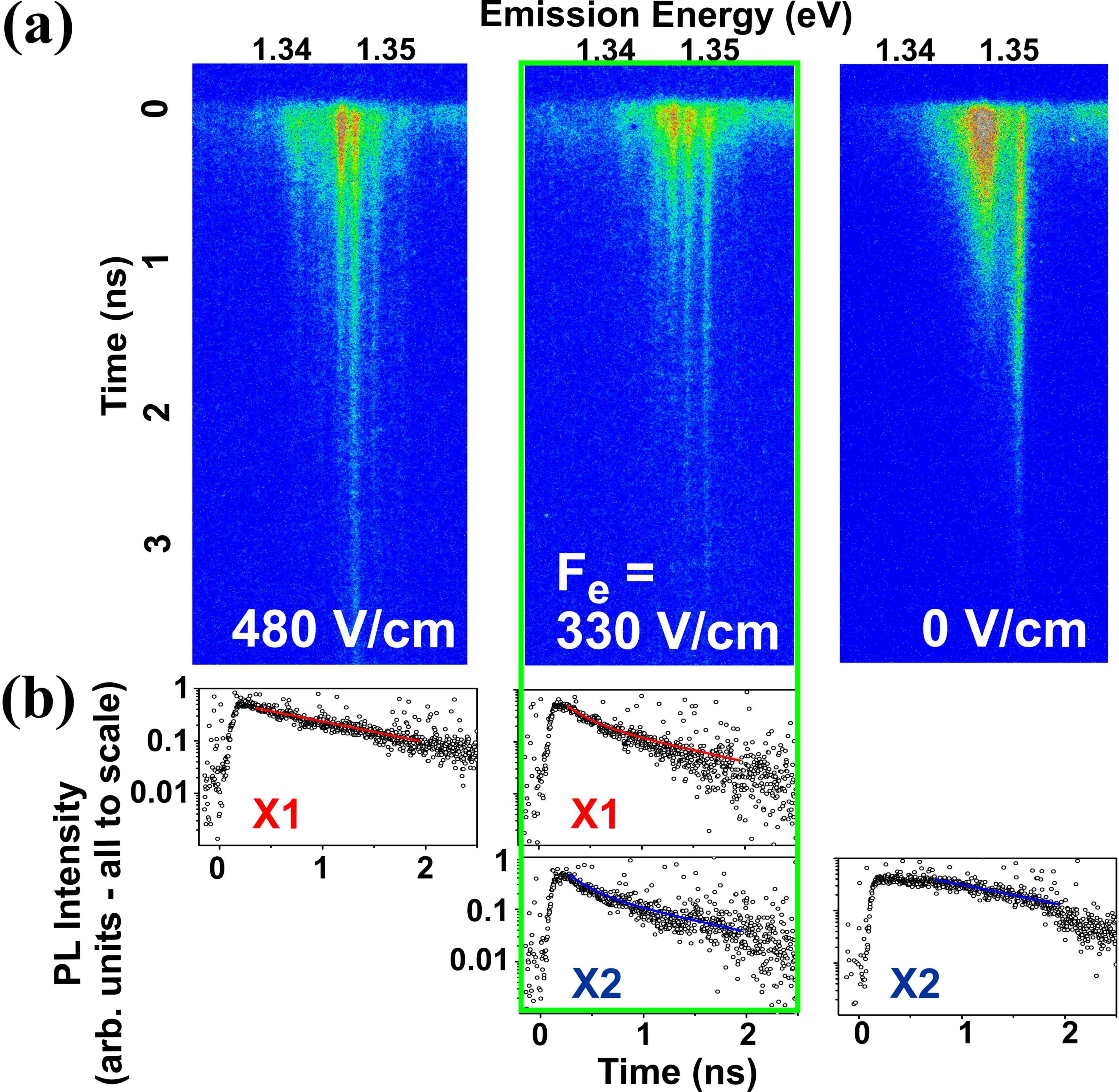}
	\caption{
\textbf{(a)} Time-energy resolved Streak camera images of the different coupling situations of the same QDM as presented in Fig.~\ref{fig:Fig1}; \textbf{(b)} the corresponding time-traces (integrated over 16 channels on the energy axis, which corresponds to a spectral range of 0.5~meV) of the excitonic recombination X1 (left panel), X1 and X2 (center panel at $F_e$) and X2 (right panel) including guides to the eye to highlight the decay behavior.
}
	\label{fig:Fig2}
\end{figure}

Due to their specific growth mode \cite{Songm1} the QDMs under investigation are all aligned along the same crystallographic axis $\left[1\bar{1}0\right]$, as displayed in the atomic force micrograph (AFM) in Fig.~\ref{fig:Fig1}~\textbf{(a)}. This alignment allows one to apply a lateral electric field using parallel electrodes on top of the sample. 
The QDMs were placed inside a low-Q planar cavity to obtain a higher photoluminescence (PL) collection efficiency and thus an enhanced signal-to-noise ratio at the typical emission energies of around 1.35~eV. This was required to simultaneously resolve the excitonic emission both spectrally and temporally. A detailed description of the structure under investigation is given in Ref. \cite{Herma2} including the demonstration of quantum coupling via photon cross-correlation spectroscopy. 
For the measurements that are presented here, the sample was mounted in a low temperature (5~K) micro-PL setup that enables the observation of single molecules within a focus diameter of about 1~$\mu$m. The QDM PL was dispersed using a 0.3~m spectrometer with a 1200~l/mm grating resulting in a theoretical Fourier transform limited spectral resolution of approximately 0.1~meV and a temporal resolution of 40~ps, which was experimentally verified by measuring a 3-ps laser pulse. The time-integrated PL spectra of a single lateral QDM under non-resonant 3-ps pulsed laser excitation were recorded using a CCD camera (Fig.~\ref{fig:Fig1}~\textbf{(b)}). The time-resolved spectra were acquired as summations of 200 background corrected 10 second exposures using a Streak camera (Fig.~\ref{fig:Fig2}). 

The effect of changing the lateral electric field on the QDM PL spectrum is displayed in Fig.~\ref{fig:Fig1}~\textbf{(b)}. Treating the charge carriers as uncorrelated single-particles, the electron states become delocalized over both dots and can be position-tuned using the electric field, whereas the hole stays almost completely localized. The two possible channels for direct neutral excitonic recombination are therefore located in one of the two coupled dots, resulting in the two recombination peaks X1 and X2. The difference in their absolute emission energy is due to small variations in dot dimensions, composition, strain, etc.
The single-particle picture is chosen here for simplicity to illustrate firstly, the possible separate treatment of the direct and indirect exciton configuration that do not mix and anticross within the investigated electric field range. The absolute energies are only lowered when excitonic correlation energies are included, which is treated in detail elsewhere \cite{Peng.Hermannstaedter.ea:2009}. Secondly, the separate tunneling of the electron and hole with significantly different rates is highlighted in this picture. This difference leads to effective transitions between indirect and direct excitons. The indirect excitonic configurations exhibit much smaller oscillator strengths due to the drastically reduced spatial overlap of electron and hole wave functions when compared to the direct case. Consequently, they are essentially optically inactive, which is in contrast to vertically coupled QDs, where the indirect excitonic recombination is, although weaker, usually visible due to a still substantial oscillator strength.

The time-resolved spectra for the direct excitons X1 and X2 are displayed in Fig.~\ref{fig:Fig2}. A clear difference is visible in the excitonic decay behavior for the three tuning situations. In the two \textit{misalignment cases}, where one of the neutral excitons dominates the spectrum (left and right panel), mono-exponential decay sets in after a slight delay (plateau), similar to a single-dot exciton in a radiative biexciton-exciton cascade. In contrast to these two cases the \emph{alignment case} at $F_e$ (central panel) exhibits a clear deviation from a single-dot like situation and reveals a double-exponential decay with a pronounced fast decay component. Comparing these situations it becomes obvious that the dynamic processes leading to the molecular tunnel-coupling clearly manifest themselves in the temporal evolution of the QDM emission.

\begin{figure}
	\centering
		\includegraphics[width=0.9\columnwidth]{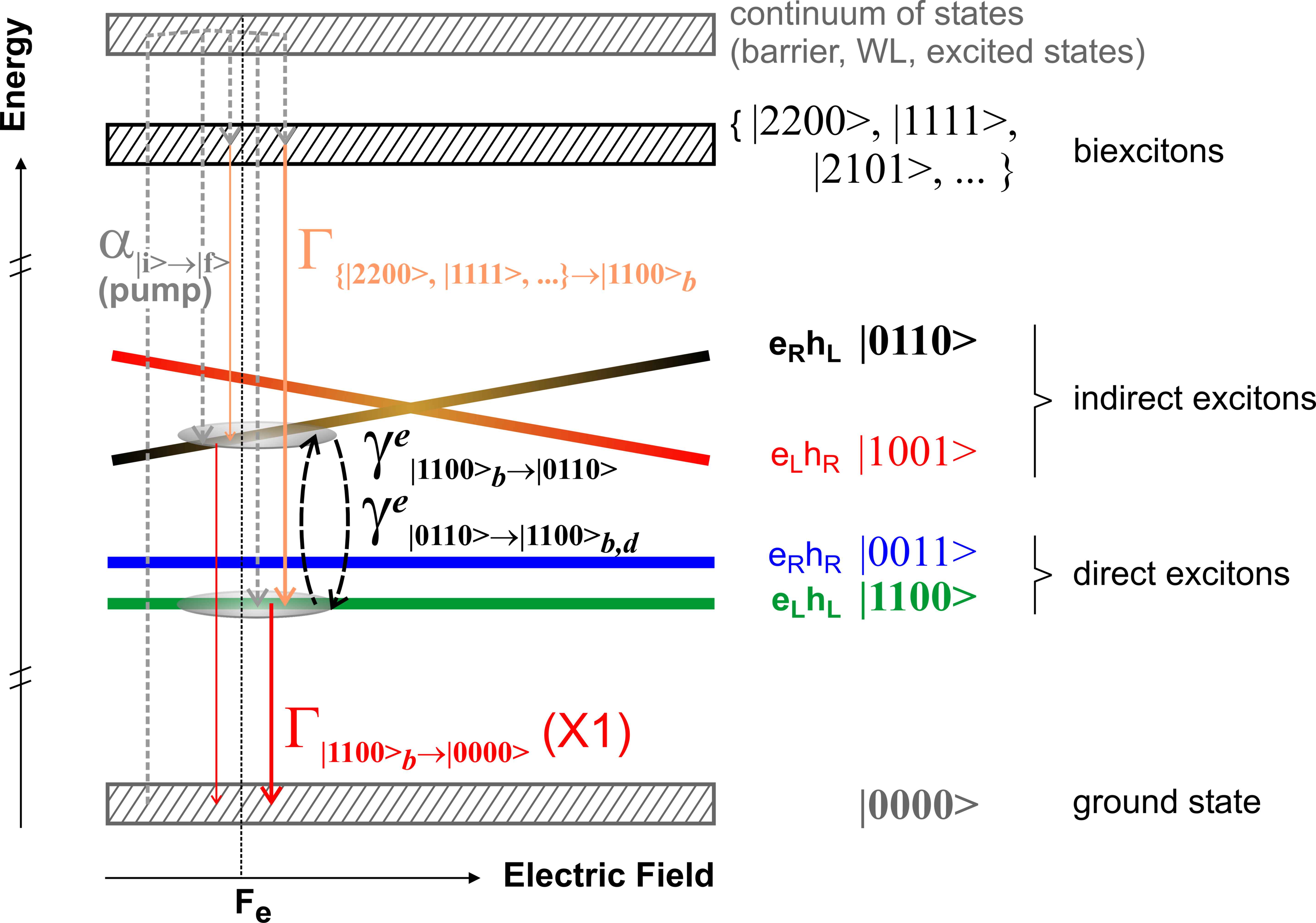}
	\caption{Schematic level diagram illustrating the molecular states, the pump, recombination and tunnel rates and the nomenclature that is used in the model.}
	\label{fig:Fig3}
\end{figure}

A rate equation model has been developed for lateral QDMs in order to describe the charge carrier dynamics and the decay behavior of the bright exciton states. The model takes account of the dynamics of excitons, radiative cascades and spin flip processes in single-dots \cite{Zwill1,Dekel1,Narva1,Reisc1}, as well as the tunneling processes reported in stacked asymmetric double-dots \cite{Villas-Boas.Govorov.ea:2004,Reisc2}. It also includes all neutral and singly charged s-shell single-particle configurations and can therefore be used to model the dynamics of excitons, charged excitons and biexcitons. The nomenclature used for the various states and rates is outlined in Fig.~\ref{fig:Fig3}.
The states are $\left|Ne_L,Nh_L,Ne_R,Nh_R\right\rangle$, with $Ne(h)=0...2$ describing the number of electrons (holes) and $L,R$ indicating the left and right dot, respectively. The carrier capture rates are $\alpha_{\left|i\right\rangle\rightarrow \left|f\right\rangle}$, the radiative rates are $\Gamma_{\left|i\right\rangle\rightarrow \left|f\right\rangle}$, where $i,f$ indicate the initial and final states of the transition. The effect of Coulomb and spin interactions on the charge carrier and exciton dynamics is included via the related tunnel rates that correspond to an effective hopping of the electron (hole), $\gamma ^{e(h)}_{\left|i\right\rangle\rightarrow \left|f\right\rangle}$, and the spin-flip rates, $\gamma ^{s}_{\left|i\right\rangle_b\rightarrow \left|f\right\rangle_d}=\gamma^{s}_{\left|i\right\rangle_d\rightarrow \left|f\right\rangle_b}$, where $b,d$ indicate the bright and dark direct exciton. 

An initially empty QDM is non-resonantly excited using a 3-ps laser pulse that is absorbed in the GaAs barrier or wetting layer (WL). The generated barrier and WL electrons, holes and excitons can either recombine at a rate of $\Gamma_{\left\{barrier,WL\right\}}\geq10$~ns$^{-1}$ or relax to the QDM states that are approximately 30--70~meV lower in energy at comparable rates of $\alpha _{\left\{barrier,WL\right\} \rightarrow \left| i \right\rangle} \geq10$~ns$^{-1}$. 
The following rate equation system was solved to describe the exciton dynamics
\begin{align}
	\frac{d}{dt} P_{\left|i\right\rangle}(t) = & -\sum_{j,k}\left\{\Gamma_{\left|i\right\rangle\rightarrow \left|j\right\rangle}\cdot P_{\left|i\right\rangle}(t) +\gamma_{\left|i\right\rangle\rightarrow \left|k\right\rangle}\cdot P_{\left|i\right\rangle}(t)\right\} \notag \\
	& +\sum_{l,m,n}\left\{\Gamma_{\left|l\right\rangle\rightarrow \left|i\right\rangle}\cdot P_{\left|l\right\rangle}(t)\left[1-P_{\left|i\right\rangle}(t)\right]\right. \notag \\
	& \left.\ \ \ +\ \alpha_{\left|m\right\rangle\rightarrow \left|i\right\rangle}\cdot P_{\left|m\right\rangle}(t) +\gamma_{\left|n\right\rangle\rightarrow \left|i\right\rangle}\cdot P_{\left|n\right\rangle}(t)\right\} , \notag
\label{eq1}
\end{align}
with the sum of all QDM states obeying the normalization condition
$\sum_{i}P_{\left|i\right\rangle}\left(t\right)=1$.
The temporal evolution of the state $\left|i\right\rangle$ is described via its population $P_{\left|i\right\rangle}(t)$. The first sum describes all processes leading to a population reduction of the state $\left|i\right\rangle$, i.e., the radiative decay processes with the rates $\Gamma_{\left|i\right\rangle\rightarrow \left|j\right\rangle}$ and the tunnel and spin-flip processes with the rates $\gamma_{\left|i\right\rangle\rightarrow \left|k\right\rangle}$. The second sum describes all processes leading to a population enhancement of the state $\left|i\right\rangle$, i.e., the radiative decay processes of the state $\left|l\right\rangle$ with the rate $\Gamma_{\left|l\right\rangle\rightarrow \left|i\right\rangle}$, e.g., in a biexciton-exciton cascade, the charge carrier capture processes with the rates $\alpha_{\left|m\right\rangle\rightarrow \left|i\right\rangle}$, and the tunnel and spin-flip processes with the rates $\gamma_{\left|n\right\rangle\rightarrow \left|i\right\rangle}$. 

The electric field tuning is included as a linear weighting factor of the capture ($\alpha$) and tunnel ($\gamma$) rates. When the QDM is in the \textit{alignment case} the carrier capture rates into the left and right dot are equal (1:1 weighting between left and right dot rates) and all direct and indirect excitons are formed, i.e., $\left|1100\right\rangle,\left|0011\right\rangle,\left|1001\right\rangle,$ and $\left|0110\right\rangle$. When the QDM is in one of the \textit{misalignment cases} the capture rates into the two dots are different (0.1:1 weighting between the corresponding rates) which corresponds to a suppression of the weaker excitonic transition of about one order of magnitude with respect to the dominant one. This ratio is in good agreement with the experimental results recorded from dozens of QDMs as a function of electric field, such as the QDM spectra presented in Fig.~\ref{fig:Fig1}~\textbf{(b)}. 
Direct exciton formation dominates in the individual dots as a result of the different rates at which the single-charge carrier capture processes take place \cite{Peng.Hermannstaedter.ea:2009}, with carrier capture and relaxation occurring at a faster rate for the electron than the hole. Initially, this leads to preferential electron occupation of either dot. This in turn enhances hole capture by the same dot due to Coulomb attraction as the exciton binding energy is about one order of magnitude larger than the opposing effect that would be expected to result from even the highest applied electric field used here.
The radiative decay rates could be obtained from experiments and are between 1.2 and 1.7~ns$^{-1}$ for the direct excitonic recombination and between 2.3 and 2.9~ns$^{-1}$ for the biexcitons. The spin-flip rate between the bright and dark excitonic configuration is assumed to be small compared to all other rates and equal in both directions \cite{Narva1}. The indirect and dark exciton recombination rates are set one order of magnitude smaller than the corresponding direct and bright exciton rates, following the argument of a reduced oscillator strength for the indirect excitons and a reduced probability of a spin-forbidden recombination \cite{Reisc1,Reisc2}. The exact values of these rates are not critical to the simulation results as long as they remain small.

The energetic separation of indirect and direct excitons leads to different tunneling rates whose magnitudes depend on whether the phonon-assisted processes require the emission or absorption of an acoustic phonon and on the phonon density of states as well as relative momenta. At the experimental low temperature conditions, phonon emission processes are possible as long as the total energy matches the acoustic phonon branch, which should be the case when $\Delta E \leq \hbar\omega^{GaAs}_{TA} \approx 9.9~meV$. Phonon absorption processes, however, depend on the thermal energy of the lattice, which consequently leads to the assumption that tunneling processes that require the absorption of a phonon must be significantly reduced. In contrast to reports on coherent tunneling in asymmetric double dots \cite{Villas-Boas.Govorov.ea:2004}, the process reported here is not coherent because it requires the coupling to acoustic phonons.

\begin{figure}
	\centering
		\includegraphics[width=0.9\columnwidth]{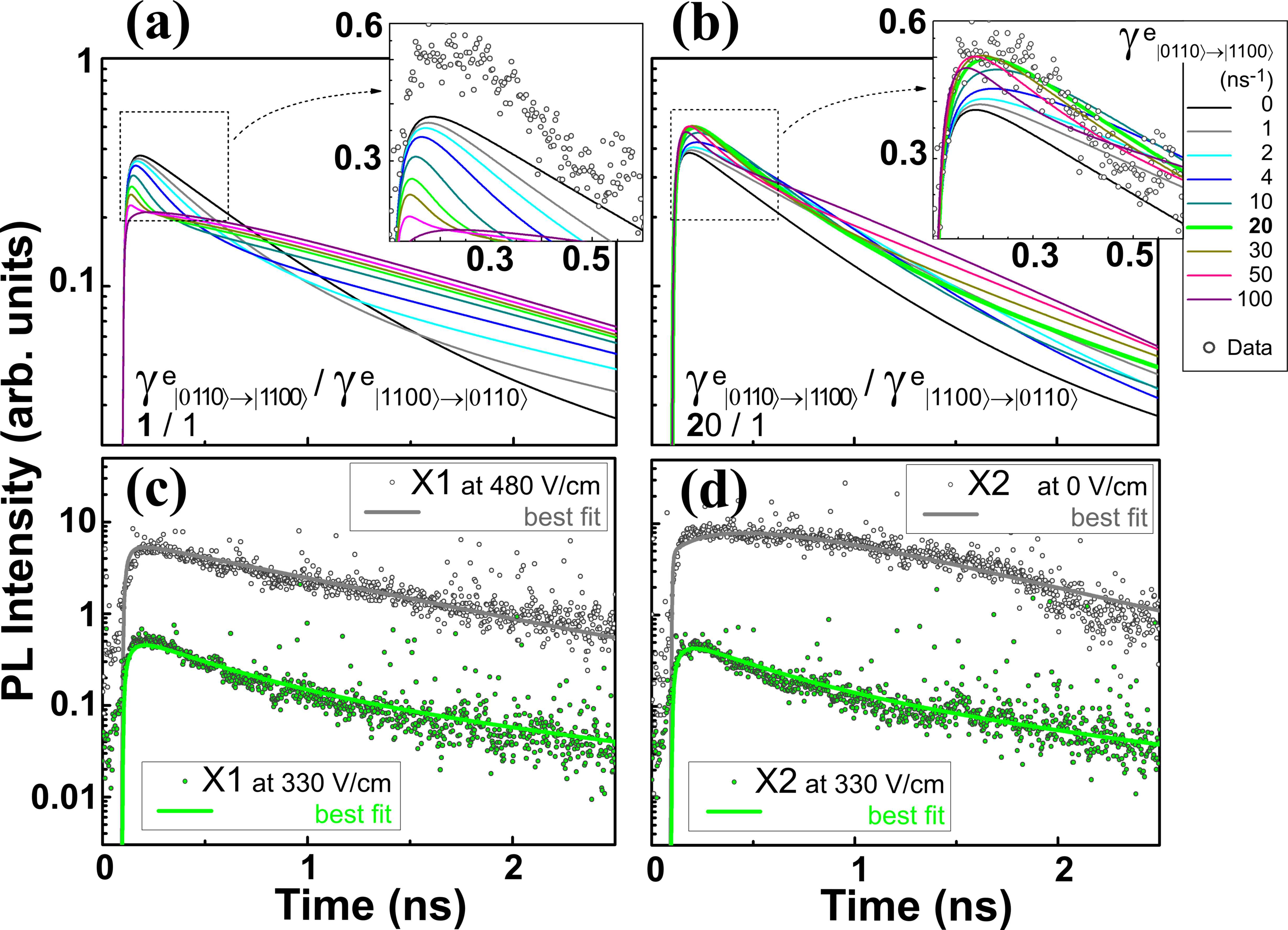}
	\caption{
The simulation results for X1 at $F_e=330$~V/cm, with the electron tunnel rate from an indirect exciton configuration varied between 0 and 100~ns$^{-1}$. The tunnel rate ratios are set as follows: \textbf{(a)} (initially indirect:direct exciton) = (1:1) and \textbf{(b)} (20:1), the values of $\gamma^e_{\left|0110\right\rangle\rightarrow \left|1100\right\rangle}$ are given as inset. 
The simulation results with best-fit parameters (lines) vs. experimental data (symbols) (the same QDM as presented in Figs.~\ref{fig:Fig1} and \ref{fig:Fig2}) for X1 \textbf{(c)} and X2 \textbf{(d)} at the different alignment conditions.
}
	\label{fig:Fig4}
\end{figure}

Fig.~\ref{fig:Fig4}~\textbf{(a,b)} compares the simulation scenarios of the X1 PL intensity, $\Gamma_{\left|1100\right\rangle_b \rightarrow \left|0000\right\rangle}\times P_{\left|1100\right\rangle_b}\left(t\right)$, where the effect of the Coulomb interactions on the tunnel rates is not included \textbf{(a)} to those where it is included \textbf{(b)} using a ratio of 20:1 between corresponding tunnel rates from indirect and direct initial configurations, e.g. $\gamma ^{e}_{\left|0110\right\rangle \rightarrow \left|1100\right\rangle}/\gamma ^{e}_{\left|1100\right\rangle \rightarrow \left|0110\right\rangle}=20$. The varied parameters are the tunnel rates; according to previous reports \cite{Wang1,Peng.Hermannstaedter.ea:2009} all tunnel rates that correspond to an effective hopping of the hole are set to small values (1~$\%$) compared to the rates that correspond to an effective hopping of the electron. Fig.~\ref{fig:Fig4}~\textbf{(a)} clearly shows that none of the simulation results fits the data due to the fast decay of the bright direct exciton population at small times (inset) and the overly slow decay at longer times, which is the result of a balanced tunneling between the different Coulomb interaction configurations (direct/indirect). Fig.~\ref{fig:Fig4}~\textbf{(b)} demonstrates that it becomes possible to fit the experimental data when Coulomb interactions are included in the simulation in terms of different tunnel rates depending on the initial state character. The best-fit cuve (green) for the X1 bright exciton recombination, $\Gamma_{\left|1100\right\rangle_b \rightarrow \left|0000\right\rangle}\times P_{\left|1100\right\rangle_b}\left(t\right)$, is obtained for the following tunnel rates: 
$\gamma ^{e}_{\left|0110\right\rangle \rightarrow \left|1100\right\rangle}=20$~ns$^{-1}$ and 
$\gamma ^{e}_{\left|1100\right\rangle \rightarrow \left|0110\right\rangle}=1$~ns$^{-1}$. 
Fig.~\ref{fig:Fig4}~\textbf{(c,d)} shows that the model can describe both the \textit{alignment} and \textit{misalignment cases}. The experimental data and simulation results obtained with the best-fit parameter set are displayed for the different electric field tuning situations. 
The X1 recombination, $\Gamma_{\left|1100\right\rangle_b \rightarrow \left|0000\right\rangle}\times P_{\left|1100\right\rangle_b}\left(t\right)$, is presented in \textbf{(c)} where the upper curve reveals the single-dot like decay below optical saturation (grey) and the lower curve presents the \textit{alignment case} (green). The X2 recombination, $\Gamma_{\left|0011\right\rangle_b \rightarrow \left|0000\right\rangle}\times P_{\left|0011\right\rangle_b}\left(t\right)$, is presented in \textbf{(d)} where the single-dot like decay case is taken at higher excitation power density leading to the onset of a plateau due to the biexciton-exciton cascaded recombination and saturation effects, which also become visible in the simulation result (grey). The lower curve presents the \textit{alignment case} (green).

It has to be noted, however, that the exciton decay behavior presented here, although very clear and pronounced, has not been observed for all of the investigated QDMs. For a particular class of molecules that exhibit their alignment at comparable and small electric fields on the order of $F_e =0\pm500$~V/cm the aforementioned effect was observed, while in some measurements performed on different QDMs with larger $\left|F_e\right|$ it was not observed. 
The reason for this could be the dominance of the quantum confined Stark effect at larger fields that somehow cancels or quenches the process leading to the decay time reduction. 
However, to-date the most likely explanation for the observed effect is the aforementioned phonon-mediated asymmetric charge carrier tunneling process, which is expected to strongly depend on the level structure of the investigated QDM and the acoustic phonon spectra. When the energy separations between correlated indirect and direct excitons does not match the phonon energies no such tunneling could occur.

In summary, we have present the charge carrier dynamics in a lateral QDM under the influence of a tuning electric field under non-resonant pulsed optical excitation. We have found that the molecular Coulomb interactions have an essential impact on the temporal evolution of the charge carrier distribution and dynamics. Using a special rate equation model that has been developed for the lateral QDM system the exciton dynamics could be modeled and the proposed asymmetric tunnel process between indirect and direct excitonic configurations could be confirmed.
We therefore suggest that non-resonant acoustic phonon mediated tunneling is the source of the observed characteristic time evolution of the presented lateral QDM excitonic spectrum. 
Such an understanding and control of the spectral and temporal properties represents a first step toward the realization of a fast optical quantum gate based on a lateral QDM.

Work at the University of Stuttgart was financially supported by the DFG via SFB/TRR 21 and FOR 730.


\begin{thebibliography}{27}
\expandafter\ifx\csname natexlab\endcsname\relax\def\natexlab#1{#1}\fi
\expandafter\ifx\csname bibnamefont\endcsname\relax
  \def\bibnamefont#1{#1}\fi
\expandafter\ifx\csname bibfnamefont\endcsname\relax
  \def\bibfnamefont#1{#1}\fi
\expandafter\ifx\csname citenamefont\endcsname\relax
  \def\citenamefont#1{#1}\fi
\expandafter\ifx\csname url\endcsname\relax
  \def\url#1{\texttt{#1}}\fi
\expandafter\ifx\csname urlprefix\endcsname\relax\def\urlprefix{URL }\fi
\providecommand{\bibinfo}[2]{#2}
\providecommand{\eprint}[2][]{\url{#2}}

\bibitem[{\citenamefont{(Ed.)}(2009)}]{Michl0}
\bibinfo{author}{\bibfnamefont{P.~Michler} \bibnamefont{(Ed.)}},
  \emph{\bibinfo{title}{Single Semiconductor Quantum Dots}}, NanoScience and
  Technology (\bibinfo{publisher}{Springer, Berlin}, \bibinfo{year}{2009}).

\bibitem[{\citenamefont{Michler et~al.}(2000)\citenamefont{Michler, Kiraz,
  Becher, Schoenfeld, Petroff, Zhang, Hu, and Imamoglu}}]{Michl1}
\bibinfo{author}{\bibfnamefont{P.}~\bibnamefont{Michler}},
  \bibnamefont{et~al.},  \bibinfo{journal}{Science} \textbf{\bibinfo{volume}{290}},
  \bibinfo{pages}{2282} (\bibinfo{year}{2000}).

\bibitem[{\citenamefont{Stevenson et~al.}(2006)\citenamefont{Stevenson, Young,
  Atkinson, Cooper, Ritchie, and Shields}}]{Steve1}
\bibinfo{author}{\bibfnamefont{R.~M.} \bibnamefont{Stevenson}},
  \bibnamefont{et~al.}, \bibinfo{journal}{Nature}, \textbf{\bibinfo{volume}{439}}, \bibinfo{pages}{179}
  (\bibinfo{year}{2006}).

\bibitem[{\citenamefont{Akopian et~al.}(2006)\citenamefont{Akopian, Lindner,
  Poem, Berlatzky, Avron, and Gershoni}}]{Akopi1}
\bibinfo{author}{\bibfnamefont{N.}~\bibnamefont{Akopian}},
  \bibnamefont{et~al.},  \bibinfo{journal}{Phys.\ Rev.\ Lett.} \textbf{\bibinfo{volume}{96}},
  \bibinfo{pages}{130501} (\bibinfo{year}{2006}).

\bibitem[{\citenamefont{Hafenbrak et~al.}(2007)\citenamefont{Hafenbrak, Ulrich,
  Michler, Wang, Rastelli, and Schmidt}}]{Hafen1}
\bibinfo{author}{\bibfnamefont{R.}~\bibnamefont{Hafenbrak}},
  \bibnamefont{et~al.},  \bibinfo{journal}{New\ J.\ Phys.} \textbf{\bibinfo{volume}{9}},
  \bibinfo{pages}{315} (\bibinfo{year}{2007}).

\bibitem[{\citenamefont{Bonadeo et~al.}(1998)\citenamefont{Bonadeo, Erland,
  Gammon, Park, Katzer, and Steel}}]{Bonad1}
\bibinfo{author}{\bibfnamefont{N.~H.} \bibnamefont{Bonadeo}},
  \bibnamefont{et~al.},  \bibinfo{journal}{Science} \textbf{\bibinfo{volume}{282}},
  \bibinfo{pages}{1473} (\bibinfo{year}{1998}).

\bibitem[{\citenamefont{Li et~al.}(2003)\citenamefont{Li, Wu, Steel, Gammon,
  Stievater, Katzer, Park, Piermarocchi, and Sham}}]{Li1}
\bibinfo{author}{\bibfnamefont{X.~Q.} \bibnamefont{Li}},
  \bibnamefont{et~al.},  \bibinfo{journal}{Science} \textbf{\bibinfo{volume}{301}},
  \bibinfo{pages}{809} (\bibinfo{year}{2003}).

\bibitem[{\citenamefont{Solomon et~al.}(1996)\citenamefont{Solomon, Trezza,
  Marshall, and Harris}}]{Solom1}
\bibinfo{author}{\bibfnamefont{G.}~\bibnamefont{Solomon}},
  \bibnamefont{et~al.},  \bibinfo{journal}{Phys.\ Rev.\ Lett.} \textbf{\bibinfo{volume}{76}},
  \bibinfo{pages}{952} (\bibinfo{year}{1996}).

\bibitem[{\citenamefont{Bayer et~al.}(2001)\citenamefont{Bayer, Hawrylak,
  Hinzer, Fafard, Korkusinski, Wasilewski, Stern, and Forchel}}]{Bayer1}
\bibinfo{author}{\bibfnamefont{M.}~\bibnamefont{Bayer}},
  \bibnamefont{et~al.},  \bibinfo{journal}{Science} \textbf{\bibinfo{volume}{291}},
  \bibinfo{pages}{451} (\bibinfo{year}{2001}).

\bibitem[{\citenamefont{Krenner et~al.}(2005)\citenamefont{Krenner, Sabathil,
  Clark, Kress, Schuh, Bichler, Abstreiter, and Finley}}]{Krenn1}
\bibinfo{author}{\bibfnamefont{H.~J.} \bibnamefont{Krenner}},
  \bibnamefont{et~al.},  \bibinfo{journal}{Phys.\ Rev.\ Lett.}
  \textbf{\bibinfo{volume}{94}}, \bibinfo{pages}{057402}
  (\bibinfo{year}{2005}).

\bibitem[{\citenamefont{Ortner et~al.}(2005)\citenamefont{Ortner, Bayer,
  Lyanda-Geller, Reinecke, Kress, Reithmaier, and Forchel}}]{Ortne1}
\bibinfo{author}{\bibfnamefont{G.}~\bibnamefont{Ortner}},
  \bibnamefont{et~al.},  \bibinfo{journal}{Phys.\ Rev.\ Lett.} \textbf{\bibinfo{volume}{94}},
  \bibinfo{pages}{157401} (\bibinfo{year}{2005}).

\bibitem[{\citenamefont{Stinaff et~al.}(2006)\citenamefont{Stinaff, Scheibner,
  Bracker, Ponomarev, Korenev, Ware, Doty, Reinecke, and Gammon}}]{Stina1}
\bibinfo{author}{\bibfnamefont{E.~A.} \bibnamefont{Stinaff}},
\bibnamefont{et~al.},  \bibinfo{journal}{Science} \textbf{\bibinfo{volume}{311}},
  \bibinfo{pages}{636} (\bibinfo{year}{2006}).

\bibitem[{\citenamefont{Krenner et~al.}(2006)\citenamefont{Krenner, Clark,
  Nakaoka, Bichler, Scheurer, Abstreiter, and Finley}}]{Krenn2}
\bibinfo{author}{\bibfnamefont{H.~J.} \bibnamefont{Krenner}},
  \bibnamefont{et~al.},  \bibinfo{journal}{Phys.\ Rev.\ Lett.}
  \textbf{\bibinfo{volume}{97}}, \bibinfo{pages}{076403}
  (\bibinfo{year}{2006}).

\bibitem[{\citenamefont{Scheibner et~al.}(2007)\citenamefont{Scheibner,
  Ponomarev, Stinaff, Doty, Bracker, Hellberg, Reinecke, and Gammon}}]{Schei1}
\bibinfo{author}{\bibfnamefont{M.}~\bibnamefont{Scheibner}},
  \bibnamefont{et~al.},  \bibinfo{journal}{Phys.\ Rev.\ Lett.} \textbf{\bibinfo{volume}{99}},
  \bibinfo{pages}{197402} (\bibinfo{year}{2007}).

\bibitem[{\citenamefont{Schmidt et~al.}(2002)\citenamefont{Schmidt, Deneke,
  Kiravittaya, Songmuang, Heidemeyer, Nakamura, Zapf-Gottwick, M{\"u}ller, and
  Jin-Phillipp}}]{Schmi1}
\bibinfo{author}{\bibfnamefont{O.~G.} \bibnamefont{Schmidt}},
  \bibnamefont{et~al.},  \bibinfo{journal}{IEEE\ J.\ Sel.\ Top.\ Quantum\
  Electron.} \textbf{\bibinfo{volume}{8}}, \bibinfo{pages}{1025}
  (\bibinfo{year}{2002}).

\bibitem[{\citenamefont{Songmuang et~al.}(2003)\citenamefont{Songmuang,
  Kiravittaya, S., and Schmidt}}]{Songm1}
\bibinfo{author}{\bibfnamefont{R.}~\bibnamefont{Songmuang}},
  \bibinfo{author}{\bibnamefont{S.}~\bibnamefont{Kiravittaya}}, \bibnamefont{and}
  \bibinfo{author}{\bibfnamefont{O.~G.}~\bibnamefont{Schmidt}},
  \bibinfo{journal}{Appl.\ Phys.\ Lett.} \textbf{\bibinfo{volume}{82}},
  \bibinfo{pages}{2892} (\bibinfo{year}{2003}).

\bibitem[{\citenamefont{Beirne et~al.}(2006)\citenamefont{Beirne,
  Hermannst{\"a}dter, Wang, Rastelli, Schmidt, and Michler}}]{Beirn1}
\bibinfo{author}{\bibfnamefont{G.~J.} \bibnamefont{Beirne}},
  \bibnamefont{et~al.},  \bibinfo{journal}{Phys.\ Rev.\ Lett.} \textbf{\bibinfo{volume}{96}},
  \bibinfo{pages}{137401} (\bibinfo{year}{2006}).

\bibitem[{\citenamefont{Robledo et~al.}(2008)\citenamefont{Robledo, Elzerman,
  Jundt, Atature, Hogele, Falt, and Imamoglu}}]{Roble1}
\bibinfo{author}{\bibfnamefont{L.}~\bibnamefont{Robledo}},
  \bibnamefont{et~al.},  \bibinfo{journal}{Science} \textbf{\bibinfo{volume}{320}},
  \bibinfo{pages}{772} (\bibinfo{year}{2008}).

\bibitem[{\citenamefont{Bester et~al.}(2004)\citenamefont{Bester, Shumway, and
  Zunger}}]{Beste1}
\bibinfo{author}{\bibfnamefont{G.}~\bibnamefont{Bester}},
  \bibinfo{author}{\bibfnamefont{J.}~\bibnamefont{Shumway}}, \bibnamefont{and}
  \bibinfo{author}{\bibfnamefont{A.}~\bibnamefont{Zunger}},
  \bibinfo{journal}{Phys. Rev. Lett.} \textbf{\bibinfo{volume}{93}},
  \bibinfo{pages}{047401} (\bibinfo{year}{2004}).

\bibitem[{\citenamefont{Wang et~al.}(2008)\citenamefont{Wang, Rastelli,
  Kiravittaya, Atkinson, Ding, Bufon, Hermannst{\"a}dter, Witzany, Beirne,
  Michler et~al.}}]{Wang1}
\bibinfo{author}{\bibfnamefont{L.}~\bibnamefont{Wang}},
  \bibnamefont{et~al.}, \bibinfo{journal}{New\ J.\ Phys.}
  \textbf{\bibinfo{volume}{10}}, \bibinfo{pages}{045010}
  (\bibinfo{year}{2008}).

\bibitem[{\citenamefont{Peng et~al.}(2009)\citenamefont{Peng, Hermannst\"adter,
  Witzany, Heldmaier, Wang, Kiravittaya, Rastelli, Schmidt, Michler, and
  Bester}}]{Peng.Hermannstaedter.ea:2009}
\bibinfo{author}{\bibfnamefont{J.}~\bibnamefont{Peng}},
  \bibnamefont{et~al.},  \bibinfo{journal}{submitted}  
   (\bibinfo{year}{2009}). \bibinfo{note}{arXiv:0910.5138v1}.

\bibitem[{\citenamefont{Hermannst\"{a}dter
  et~al.}(2009)\citenamefont{Hermannst\"{a}dter, Witzany, Beirne, Schulz,
  Eichfelder, Rossbach, Jetter, Michler, Wang, Rastelli et~al.}}]{Herma2}
\bibinfo{author}{\bibfnamefont{C.}~\bibnamefont{Hermannst\"{a}dter}},
  \bibnamefont{et~al.}, \bibinfo{journal}{J. Appl. Phys.}
  \textbf{\bibinfo{volume}{105}}, \bibinfo{eid}{122408} (\bibinfo{year}{2009}).

\bibitem[{\citenamefont{Zwiller et~al.}(1999)\citenamefont{Zwiller, Pistol,
  Hessman, Cederstr\"om, Seifert, and Samuelson}}]{Zwill1}
\bibinfo{author}{\bibfnamefont{V.}~\bibnamefont{Zwiller}},
  \bibnamefont{et~al.},  \bibinfo{journal}{Phys. Rev. B} \textbf{\bibinfo{volume}{59}},
  \bibinfo{pages}{5021} (\bibinfo{year}{1999}).

\bibitem[{\citenamefont{Dekel et~al.}(2000)\citenamefont{Dekel, Regelman,
  Gershoni, Ehrenfreund, Schoenfeld, and Petroff}}]{Dekel1}
\bibinfo{author}{\bibfnamefont{E.}~\bibnamefont{Dekel}},
  \bibnamefont{et~al.},  \bibinfo{journal}{Phys. Rev. B}
  \textbf{\bibinfo{volume}{62}}, \bibinfo{pages}{11038} (\bibinfo{year}{2000}).

\bibitem[{\citenamefont{Narvaez et~al.}(2006)\citenamefont{Narvaez, Bester,
  Franceschetti, and Zunger}}]{Narva1}
\bibinfo{author}{\bibfnamefont{G.~A.} \bibnamefont{Narvaez}},
  \bibnamefont{et~al.},  \bibinfo{journal}{Phys.\ Rev.\ B} \textbf{\bibinfo{volume}{74}},
  \bibinfo{pages}{205422} (\bibinfo{year}{2006}).

\bibitem[{\citenamefont{Reischle et~al.}(2008)\citenamefont{Reischle, Beirne,
  Ro{\ss}bach, Jetter, and Michler}}]{Reisc1}
\bibinfo{author}{\bibfnamefont{M.}~\bibnamefont{Reischle}},
  \bibnamefont{et~al.},  \bibinfo{journal}{Phys. Rev. Lett.} \textbf{\bibinfo{volume}{101}},
  \bibinfo{pages}{146402} (\bibinfo{year}{2008}).

\bibitem[{\citenamefont{Villas-B\^oas et~al.}(2004)\citenamefont{Villas-B\^oas,
  Govorov, and Ulloa}}]{Villas-Boas.Govorov.ea:2004}
\bibinfo{author}{\bibfnamefont{J.~M.} \bibnamefont{Villas-B\^oas}},
  \bibinfo{author}{\bibfnamefont{A.~O.} \bibnamefont{Govorov}},
  \bibnamefont{and} \bibinfo{author}{\bibfnamefont{S.~E.} \bibnamefont{Ulloa}},
  \bibinfo{journal}{Phys. Rev. B} \textbf{\bibinfo{volume}{69}},
  \bibinfo{pages}{125342} (\bibinfo{year}{2004}).

\bibitem[{\citenamefont{Reischle et~al.}(2007)\citenamefont{Reischle, Beirne,
  Ro{\ss}bach, Jetter, Schweizer, and Michler}}]{Reisc2}
\bibinfo{author}{\bibfnamefont{M.}~\bibnamefont{Reischle}},
  \bibnamefont{et~al.},  \bibinfo{journal}{Phys. Rev. B} \textbf{\bibinfo{volume}{76}}
  (\bibinfo{year}{2007}).

\end{thebibliography}

\end{document}